\documentstyle[epsf,preprint,eqsecnum,aps]{revtex}
\begin{document}
\draft
\preprint{Alberta-Thy-06-95/ McGill 95--11}
\date{March 1995}

\title{ON SOME NEW BLACK STRING SOLUTIONS IN THREE DIMENSIONS}
\author{
    Warren G. Anderson$^{1}$\footnote{anderson@phys.ualberta.ca} and
    Nemanja Kaloper$^{2}$\footnote{kaloper@hep.physics.mcgill.ca}}
\address{
  $^1$Theoretical Physics Institute, University of  Alberta,\\
      Edmonton, Alberta, Canada T6G 2J1\\
  $^2$Department of Physics, McGill University, \\
      Montreal, Quebec, Canada H2A 2T8.}
\maketitle
%
%

\begin{abstract}
We derive several new
solutions in three-dimensional
stringy gravity. The solutions are obtained with
the help of string duality transformations.
They represent stationary configurations with
horizons, and are surrounded by
(quasi) topologically massive Abelian gauge hair,
in addition to the dilaton and the Kalb-Ramond axion.
Our analysis suggests that there exists a more general family,
where our solutions are special limits. Finally, we use the
generating technique recently proposed
by Garfinkle to construct a traveling wave on the
extremal variant of one of our solutions.
\end{abstract}
\pacs{PACS numbers: 04.20.Jb, 04.50.+h, 12.10.Gq, 97.60.Lf\\
{\it Submitted to Phys.  Rev. D}}
\clearpage

\newcommand{\be}{\begin{equation}}
\newcommand{\ee}{\end{equation}}
\newcommand{\bea}{\begin{eqnarray}}
\newcommand{\eea}{\end{eqnarray}}
\newcommand{\di}{\partial}
\newcommand{\sh}{\hat{s}}
\newcommand{\ch}{\hat{c}}
\newcommand{\rh}{\hat{r}}
\newcommand{\rb}{\bar{r}}
\newcommand{\rhoh}{\hat{\rho}}
\newcommand{\xih}{\hat{\xi}}
\newcommand{\tauh}{\hat{\tau}}

\section{Introduction}
Recently we have witnessed a surge of interest
in lower-dimensional theories of gravity, after the realization that many of
them contain structures with horizons \cite{EW}-\cite{sols}.
Investigation of these models is
motivated by the hope that we may be able to gain more information
about the physics of realistic four-dimensional black holes, since
mathematical difficulties subside dramatically in fewer dimensions.
This approach appears to be particularly fruitful in lower-dimensional stringy
gravity, where the facilities of string
theory provide us with very powerful
tools to study black holes in the classical limit and beyond
\cite{ODD}-\cite{KUM}.
Resorting to these techniques, we may be able to tackle in a systematic way
some of the long-standing conundrums of black hole physics.
For example, it has been demonstrated that string theory may have the
potential to cure some of the singularity problems which plague the classical
theory\footnote{This should be taken with some reservations, because the
specific conclusions obtained so far
may not hold in general, as shown recently
by Horowitz and Tseytlin \cite{HorTse}.}\cite{PTY}.

In this paper we shall contribute several new black hole-like solutions
to the existing bestiary. We shall employ the Abelian duality symmetry
as our main tool to obtain them \cite{ODD}-\cite{CFT}.
Such symmetries represent a
stringy generalization of standard toroidal
symmetries, stemming from the presence of commuting
translational Killing vectors in a gravitational background.
They can be combined and  employed to derive
new background solutions. The general procedure
is dubbed twisting, or $O(d,d)$ boosting, after the
complete group of twisting transformations \cite{VEN1,SEN1,MS}.
At the level of the background field theory on target space,
after integrating out Killing coordinates \`a la
Kaluza-Klein, this is realized as a
symmetry of the action under mixing of the Kaluza-Klein matter
fields with the metric.
Although the action is invariant
under this group, solutions are not,
because they employ specific initial conditions.
Therefore, the twisting transformations
can generate new classical solutions. It must be kept in mind,
however, that dual solutions may not represent different string
physics, but merely be different pictures of the same string theory,
which for example occurs when one dualizes with respect to the translation
of a compact coordinate \cite{RV}.
Moreover, the full $O(d,d)$ group also contains
diffeomorphisms and Kalb-Ramond field gauge
transformations, which must be modded out \cite{SEN1}.
Thus, the space of classical solutions is spanned by the
orbits of the $O(d,d)$ group, modulo diffeomorphisms
and Kalb-Ramond gauge transformations.

This symmetry is further extended to $O(d,d+n)$
in the presence of $n$ Abelian
gauge fields \cite{MSS}.
We will use this extended boost symmetry to obtain two new
three-dimensional (3D) families of asymptotically flat solutions.
Our families are obtained by, respectively,
``twisting in" the gauge field on
the black string of Horne and Horowitz \cite{HorHo},
and the axion on the 2D electrically charged black hole crossed with a
flat line \cite{NMY}. They are characterized
by three parameters, and for certain ranges of the parameters
they represent different stationary, gauge
charged configurations with regular horizons.
Some of their properties are quite remarkable.
Namely, although all our non-extremal black strings
possess a scalar curvature singularity, which must be
included in the manifold because it is spacelike geodesically incomplete,
this singularity is quite harmless for pointlike observers. The manifold is
null and timelike geodesically complete, since for arbitrary
initial conditions at infinity all causal geodesics have a turning point
before reaching the singularity except for one null geodesic,
which comes arbitrarily
close to the singularity but never reaches it for a finite
value of the affine parameter. The first family
possesses an interesting non-singular extremal limit, different from the
extremal limits of previously known solutions in that it has one
hypersurface orthogonal null Killing vector, and nonvanishing gauge hair.
Therefore, we can further extend this solution to
include a traveling wave,
using the generating technique proposed recently by D. Garfinkle in
the context of string gravity \cite{Gar}.
Our first family also contains a subclass of solutions
with interesting global properties. These solutions are
without curvature singularities, with spatial hypersurfaces
looking like a ``cigar'', approaching asymptotically
${\bf R} \times {\bf S}^1$.
However, the angular variable is ``bolted'' to time,
and hence near the origin there appear closed timelike curves.

The second family has, curiously, a critical value of the boost
parameter and a black string with two different extremal limits.
The critical boost gives the stringy version of the
3D black hole \cite{BTZ,HWK}. Away from this critical boost we find another
family of black strings, which displays a peculiar combination
of properties of both black strings and four dimensional
black holes. Particularly interesting is the presence of the
ergosphere, which arises entirely due to the axion and electric
charges. Furthermore, this black string has two extremal limits.
One of them corresponds to taking the critical value of the boost,
but after a coordinate transformation, in a fashion familiar from
the static black string case. The other extremal limit is reminiscent
of the extremal Kerr-Newman case, and represents a
gauge charged black string, with ergosphere but without null Killing vectors.

The paper is organized as follows. In the next section, we will lay
out mathematical background for the subsequent study, explaining
our approach and deriving various forms of the solutions. Detailed
investigation of the solutions will be presented in section III.
Section IV contains the derivation of the traveling wave solution,
using Garfinkle's techniques.
The final section offers our conclusions and presents arguments
that suggest the existence of
a larger family of 3D black objects which continually
interpolate between our two solutions, as well as the Horne-Horowitz and
the stringy BTZ solutions.

\section{Generating Solutions}
The effective action of string theory describing dynamics of
massless bosonic background fields to the lowest
order in the inverse string tension $\alpha'$ is, in the
world-sheet frame \cite{VEN1}-\cite{MS},\cite{GS},
\begin{equation}\label{w1}
S~=~\int d^{d+1}x\sqrt{g}  e^{- \Phi}
\big(R +\partial_{\mu}\Phi \partial^{\mu} \Phi
-{1 \over 12} H_{\mu\nu\lambda}H^{\mu\nu\lambda}
-{\alpha' \over 4} F^{N}{}_{\mu\nu}F^{N}{}^{\mu\nu}
+ 2 \Lambda \big).
\end{equation}
The action above is written in Planck units $\kappa^2 = 1$.
Here $~F^{N}{}_{\mu\nu}= \partial_{\mu}A^{N}{}_{\nu} -
\partial_{\nu}A^{N}{}_{\mu}~$
are field strengths of $n$ Abelian gauge fields $A^{j}{}_{\mu}$,
$~H_{\mu\nu\lambda}= \partial_{\lambda}B_{\mu\nu} +
{}~cyclic~permutations~
- ({\alpha'/2}) \Omega_M{}_{\mu\nu\lambda}~$
is the field strength associated with the Kalb-Ramond
field $~B_{\mu\nu}~$ and $~\Phi~$
is the dilaton field. The Maxwell Chern-Simons form
$\Omega_M{}_{\mu\nu\lambda} =
\sum_N A^{N}{}_{\mu}F^{N}{}_{\nu\lambda}
+ ~cyclic~permutations~ $ appears
in the definition of the axion field strength due to the
Green-Schwarz anomaly
cancellation mechanism, and can be understood as a model-independent
residue after dimensional reduction from ten-dimensional superstring
theory \cite{WDim}. In fact, this term is a necessary
ingredient of the theory if one wants to ensure the
$O(d,d+n)$ invariance, as shown by Maharana and Schwarz \cite{MS}.
The $n$ Abelian gauge fields should be thought of as
the components of a non-Abelian gauge field ${\bf A}$ residing in the
Cartan subalgebra of the gauge group, while the
rest have been set equal to zero.
For convenience we will set $\alpha' = 1$.

In what follows we will be considering only those extrema of
(\ref{w1}) that possess
$d$ commuting isometries, that is, we will be considering field
configurations of the
form
\begin{eqnarray}\label{w2}
ds^{2} &=& \Gamma(r) ~dr^{2}+G_{jk}(r)~dx^j dx^k, \nonumber \\
B &=& \frac{1}{2}~ B_{jk}(r) dx^j \wedge dx^k, \\
A^{N} &=& A^{N}{}_{j}(r) dx^j,  \nonumber \\
\Phi &=& F(r),  \nonumber
\end{eqnarray}
where $G_{jk}(r)$ is the metric of a $d$ dimensional submanifold of
signature $d-2$.  In this case the action (\ref{w1}) can be rewritten in the
manifestly $O(d,d+n)$ invariant form \cite{VEN1,SEN1,MS}
\begin{equation}\label{w3}
S_{eff}~=~\int d r \sqrt{\Gamma}  e^{-\phi}
\Big({1 \over \Gamma}\phi'^2
+ {1 \over 8\Gamma}\mbox{Tr}\bigl({\cal M}'{\cal L}\bigr)^2 + 2
\Lambda\Big),
\end{equation}
where the prime denotes the derivative with respect to $r$. Note that
the physical
dilaton $\Phi$ has been replaced by the effective dilaton
$\phi = \Phi - (1/2) \ln |\det G |$ after dimensional reduction.
Matrices
$~{\cal M}~$ and $~{\cal L}~$  which appear in the action (\ref{w3})
are defined
by
\begin{eqnarray}\label{w4}
&& {\cal M}~=~\pmatrix{
g^{-1}&-g^{-1}C&-g^{-1}A \cr
-C^{T}g^{-1}&g  + a + C^{T}g^{-1}C&A + C^{T}g^{-1}A \cr
-g^{-1}A&A + C^{T}g^{-1}A&{\bf 1}_n +  A^{T}g^{-1}A \cr} ,
\nonumber\\
&&~~~~~~~~  \\
&&~~~~~~~~~~~~~~~~~~~ {\cal L}~=~\pmatrix{~0~&{\bf 1}_d&~0~ \cr
{\bf 1}_d&~0~&~0~ \cr
{}~0~&~0~&{\bf 1}_n \cr}. \nonumber
\end{eqnarray}
Here $~g~$ and $~b~$ are $~d \times d~$ matrices
defined by the dynamical degrees of freedom of the metric and the
axion:$~g~=~\bigl(G_{jk}\bigr)$ and $b~=~\bigl(B_{jk}\bigr)$. The
matrix $A$ is
a $~d \times n~$ matrix built out of the gauge fields:
$A_{kN}~=~A^{N}{}_{k}$. The matrices $a$  and $C$ are defined by
$a =  AA^{T}$ and $C = (1/2)a + b$ respectively, and $~{\bf 1}_d~$
and $~{\bf 1}_n~$ are the $d$ and $n$ dimensional unit matrices. Note
that ${\cal M}^{T} = {\cal M}$ and
${\cal M}^{-1} = {\cal L}{\cal M}{\cal L}$. Thus we see that ${\cal M}$ is a
symmetric element of $O(d,d+n)$. Therefore a cogradient $O(d,d+n)$
rotation ${\cal M} \rightarrow \Omega {\cal M} \Omega^{T}$ is a symmetry
of the action, and the equations of motion, because it represents a group
motion which changes $\cal M$ while maintaining its symmetry property.

In this paper we will apply this technique to several well-known
solutions of stringy gravity to the lowest order in the inverse
string tension expansion, describing black strings. Specifically,
we will use the black string family
discovered by Horne and Horowitz \cite{HorHo},
as well as the 2D electrically charged black
hole crossed with a flat line, discussed
by McGuigan et al. \cite{NMY}.
We start with the black string solution, given by
\cite{HorHo}:
\bea \label{w7}
     ds^2 = \frac{dr^2}{2\Lambda (r-m)(r-\frac{Q^2}{m})} &+&
         (1-\frac{Q^2}{mr}) dx^2 - (1-\frac{m}{r}) dt^2, \nonumber \\
     B = \frac{Q}{r} dx \wedge dt, &&~~~~~~~~ A=0, \\
     e^{- \Phi} &=& \sqrt{2\Lambda} r \nonumber.
\eea
Following the prescription outlined above, and using
\be \label{w9}
     \Omega = \left (\matrix{
          1 & 0 & 0 & 0 & 0 \cr
          0 & (1+c)/2 & 0 & (1-c)/2 & -s/\sqrt{2} \cr
          0 & 0 & 1 & 0 & 0 \cr
          0 & (1-c)/2 & 0 & (1+c)/2 & s/\sqrt{2} \cr
          0 & -s/\sqrt{2} & 0 & s/\sqrt{2} & c \cr
          }\right ),
\ee
where $c = \cosh(2\alpha)$ and $s = -\sinh(2\alpha)$,
we obtain the new solution
\bea \label{w10}
     ds^2 &=& \frac{d\rh^2}{2\Lambda
(\rh-m\ch^2)(\rh-m\sh^2-Q^2/m)} +
          \frac{\rh-m\sh^2-Q^2/m}{\rh-m\sh^2} dx^2 ~~~~~~ \nonumber \\
&&~~~~~~~~~~~~~~~~~~~~~~~~~~~~~~~- \frac{\rh-m\ch^2}{\rh^2(\rh-m\sh^2)}
     ((\rh-m\sh^2) dt - \sh^2 Q dx)^2, \nonumber \\
    &&~~~ B = \frac{Q \ch^2}{\rh} dx \wedge dt, ~~~~~~~~~~~~~
A=-\sqrt{2}\frac{\sh\ch}{\rh}(Q dx + m dt), \\
    &&~~~~~~~~~~~~~~~~~~~~~~ e^{-\Phi} = \sqrt{2 \Lambda} \rh, \nonumber
\eea
where $\ch = \cosh \alpha$, $\sh = -\sinh \alpha$, and
$\rh = r + m \sh^2$. Clearly, (\ref{w10}) generalizes (\ref{w7}),
reducing to it when $\alpha = 0$. It is obvious that this
solution, like (\ref{w7}), is asymptotically flat in the limit
$\rh \rightarrow \infty$.
We also find it useful to represent our
new solution in terms of the shifted time coordinate
$\tau = t + (Q/m)x$ instead of $t$.
In this gauge, the axion and the dilaton are the same as above. The
gauge field is oriented completely along $d\tau$, and
the expressions for it and the metric are given as follows:
\bea \label{w10spec}
     ds^2 &=& \frac{d\rh^2}{2\Lambda
(\rh-m\ch^2)(\rh-m\sh^2-Q^2/m)} +
          \bigl(1 - \frac{Q^2}{m^2}\bigr) dx^2 ~~~~~~~~~ \nonumber \\
&&~~~~~~~~~~~~~~~~~~- 2\frac{Q}{m} \bigl(1 - \frac{m\ch^2}{\rh}\bigr) dxd\tau
- \bigl(1 - \frac{m\ch^2}{\rh}\bigr)
\bigl(1 - \frac{m\sh^2}{\rh}\bigr) d\tau^2, \\
&&~~~~~~~~~~~~~~~~~~~~~~~ A=-\sqrt{2}\frac{m\sh\ch}{\rh}d\tau. \nonumber
\eea
This form is suitable for comparison with our
second solution, to be presented shortly, but not useful for the
analysis of the causal structure, as can be immediately seen from the
fact that $\tau$ is not the asymptotic time coordinate.

We now consider the electrically charged stringy 2D black
hole crossed with a flat line ${\bf R}$ \cite{NMY}:
\bea \label{w13}
   ds^2=\frac{d{\rho}^2}{2\Lambda
({\rho}-\frac{q^2}{\mu})({\rho}-\mu)} +&d\xi^2&-
      (1-\frac{q^2}{\mu\rho})
   (1-\frac{\mu}{\rho})d\hat\tau^2, \nonumber \\
   B =0, ~~~~~~~~~~~&& A= -\frac{\sqrt{2} q}{\rho} d\hat\tau, \\
   e^{-\Phi}=&\sqrt{2\Lambda}&{\rho}, \nonumber
\eea
To obtain our other solution we apply
another $O(2,3)$ twist to it, with $b$ a real number:
\be \label{w14}
   \Omega = \left (\matrix{
      0 & 1 & b & 0 & 0 \cr
      1 & 0 & 0 & -b & 0 \cr
      0 & 0 & 0 & 1 & 0 \cr
      0 & 0 & 1 & 0 & 0 \cr
      0 & 0 & 0 & 0 & 1 \cr
   }\right ).
\ee
This yields the following expression:
\bea \label{w16}
ds^2 &=&
\frac{d{\rho}^2}{2\Lambda({\rho}-\frac{q^2}{\mu})({\rho}-\mu)}
+ \frac{b^2 ({\rho}-M)^2-f}{((1-b^2){\rho}+Mb^2)^2}
d\xi^2  \nonumber \\
&& ~~~~~~~~~~~~~~~
+ \frac{{\rho}^2-b^2f}{((1-b^2){\rho}+Mb^2)^2} d\hat\tau^2
- \frac{2 b q^2}{((1-b^2){\rho}+Mb^2)^2} d\xi d\hat\tau, \nonumber \\
   B &=& \frac{b({\rho}-M)}{(1-b^2){\rho}+Mb^2} d\xi
\wedge d\hat\tau,
   ~~~~~A = - \frac{\sqrt{2} q}{(1-b^2){\rho}+Mb^2}(d\xi + b
d\hat\tau),  \\
&&~~~~~~~~~~~~~ e^{-\Phi}= \sqrt{2 \Lambda} ((1-b^2){\rho}+Mb^2),
\nonumber
\eea
where $f=(\rho - \mu)(\rho - q^2/\mu)$ and $M=\mu+\frac{q^2}{\mu}$.
Note that $b^2 = 1$ represents a special point in the moduli
space of this family, as the dilaton field decouples there.
Indeed, after a closer look (and some coordinate transformations)
we recognize this case as precisely
the stringy version of the Banados-Teitelboim-Zanelli (BTZ) \cite{BTZ,HWK}
black hole. In what follows, for computational purposes we will
assume $b > 1$ (the sign of $b$ only determines the sign of the axion charge),
without any loss of generality, as we will now demonstrate.
To start with, the solution (\ref{w16}) can be simplified
considerably with some judicious gauge choices. We begin with the
coordinate
transformation $\rh = (1-b^2){\rho}+Mb^2$, $x= \xi/\sqrt{b^2-1}$,
and $t=\hat\tau/\sqrt{b^2-1}$.
We can also apply the axion gauge
transformation which shifts the asymptotic value
of the axion to zero. In this gauge, the solution takes the form
\bea \label{w17}
ds^2 &=&
\frac{d\rh^2}{2\Lambda(\rh-\mu b^2-\frac{q^2}{\mu})
(\rh-\mu-\frac{q^2b^2}{\mu})} +
(1-\frac{M}{\rh}-\frac{q^2(b^2-1)}{\rh^2}) dx^2 ~~~~~~~ \nonumber \\
&&~~~~~~~~~~~~~~~~~~- (1-\frac{Mb^2}{\rh}
+\frac{q^2(b^2-1)b^2}{\rh^2})dt^2 -
     2 \frac{b q^2(b^2-1)}{\rh^2} dxdt \nonumber \\
   &&~ B = \frac{bM}{\rh} dx \wedge dt,
   ~~~~~~~~~~~~~A = - \sqrt{2}\frac{q\sqrt{b^2-1}}{\rh}(dx + b dt),  \\
 &&~~~~~~~~~~~~~~~~~~~~~~~  e^{-\Phi}= \sqrt{2 \Lambda} \rh. \nonumber
\eea
It is straightforward to verify that if $|b|<1$, the same coordinate
transformation, followed by Wick rotation $t,x \rightarrow it, ix$,
rescaling $\rh \rightarrow \rh/b^2$, a constant dilaton shift and
the replacement of the parameter
$b \rightarrow b' = 1/b >1$ reduces the form of (\ref{w16})
again to (\ref{w17}). We will defer further discussion of the
interpretation of the $|b|<1$ solutions until later.

The solution (\ref{w17}) is also an asymptotically flat configuration with
both axion and gauge fields. It is again useful, for easier comparison,
to perform another coordinate
change, to put this solution in a form similar to (\ref{w10spec}). The
dilaton and axion remain the same as above, while the metric and the
gauge field are given as follows:
\bea \label{w17spec}
ds^2 &=&
\frac{d{\rh}^2}{2\Lambda({\rh}-\mu b^2-\frac{q^2}{\mu})
({\rh}-\mu -\frac{q^2b^2}{\mu})} + \bigl(1 - \frac{1}{b^2}\bigr) dx^2
{}~~~~~~ \nonumber \\
&&~~~~~~~~~~~~~~ - 2 \frac{1}{b}\bigl(1-\frac{Mb^2}{\rh}\bigr) dxd\tau
- (1-\frac{Mb^2}{\rh}
+\frac{q^2(b^2-1)b^2}{\rh^2})d\tau^2, \\
&&~~~~~~~~~~~~~~~~~~~~ A= - \sqrt{2}\frac{qb\sqrt{b^2-1}}{\rh}d\tau. \nonumber
\eea
Despite the conspicuous similarity between (\ref{w10spec}) and
(\ref{w17spec}), we will demonstrate later in our analysis that they are
indeed different. This can already be glimpsed, however, by
realizing that the matter content of the two configurations is
exactly the same once the proper coordinate rescalings
are performed, and that since they are
stationary and contain a scalar, a vector and the volume form
in the $(x,\tau)$ subspace, we are left without any freedom
to perform further coordinate transformations which do not alter the
form of the matter. Specifically, the fact that the dilaton of both
configurations is essentially the radial coordinate, and that $x$ and
$\tau$ are Killing coordinates restricts the available coordinate
transformations to only linear transformations in the $(x,\tau)$ plane.
These in general induce the changes of the two gauge fields $A$ and $B$
which are proportional to the field components themselves. Since the
fields are nontrivial, i.e. have non-vanishing field strengths, the
changes induced by diffeomorphisms are not pure gauges and hence cannot
be removed by gauge transformations. The last step in the argument is
the comparison of the two metrics, which shows that
they do not match; indeed, if we denote the $(x,\tau)$ parts of the two
metrics as $g_1$ and $g_2$, respectively, we can see that
$g_1 = g_2 + Cdx^2$, for some given constant $C$. Since the horizons are
determined by the determinant of these matrices, the above shift
induces the corresponding shift in the locations of these surfaces.

In the next section we will investigate causal properties of
these solutions. Our analysis will confirm and elaborate upon the
argument presented above, that they represent
different black strings. Before we close this section, however,
we should explain an apparent peculiarity which appears in the gauge
sector of the two solutions. Namely, we see that in both
(\ref{w10spec}) and (\ref{w17spec}) the gauge field looks
precisely like the field of a point charge in three spatial dimensions,
despite the fact that it lives in two dimensions, where one would
expect it to be proportional to the logarithm of the distance from the
source. Indeed, such behavior has been noted in Ref. \cite{BTZ},
where charged black holes in 3D Einstein-Maxwell theory were studied.
The resolution to this lies in the fact that
in our background the gauge field acquires the (quasi)topological mass
term due to its coupling to the Kalb-Ramond field via the Chern-Simons
form \cite{DJT,Kaltmg}. The Kalb-Ramond field is
trivially integrable in three dimensions \cite{Kaltmg},
and if non-zero, yields the gauge field mass term.
The standard Maxwell equation for the gauge field should be
replaced in this case by, in form notation,
$d \exp(-\Phi) {^{*} {\bf F}} = 2Q_A {\bf F}$,
where $Q_A$ is the axion charge,
defined by $Q_A = -\exp(-\Phi) {^{*}{\bf H}}$;
it is straightforward to verify that our backgrounds solve it.
It is furthermore interesting to note, that whereas in this case
the gauge field is (quasi)topologically massive,
the gauge sector of either solution does not represent
a gauge anyon, as the Chern-Simons form itself vanishes in both cases.

\section{Causal Structure}
Here we will investigate the structure of the two new
solutions presented above. Whereas some aspects of the
geometry of these two solutions are remarkably similar,
there are interesting differences.

As a warm-up, let us review the static black string (\ref{w7}) of
Horne and Horowitz,
which will be the basis for comparison. This solution has three metric
singularities at $r=m$, $r=Q^2/m$ and $r=0$, and obviously there are
three different cases, $0<|Q|<m$, $|Q|=m$, and $|Q|>m$. When
$0<|Q|<m$, $r=0$ is a scalar curvature singularity and
$r=m$ and $r=Q^2/m$ are the event and Cauchy
horizons respectively. The singularity is ``real" in the sense that
the manifold is null geodesically incomplete.
The causal structure of the solution is qualitatively similar
to that of the Reissner-Nordstr$\o$m solution, with the
exception that the time-like coordinate inside
the Cauchy horizon is $x$ rather than $t$. Thus the 2D
Penrose diagrams aren't completely adequate for the description of
the geometry, but they can be used with the proviso that one remembers that
the time-like coordinate makes a ``right angle" turn on the inner horizon.
As a consequence, in this
solution there are no static observers inside the Cauchy horizon.
This case is summarized in Fig.1.

When $|Q|=m$, the form of the solution (\ref{w7}) breaks down at the
(degenerate) horizon $r=m$. It turns out that the coordinate $r$ is not
suitable for the extension beyond the event horizon, which appears to
be a turning point for all geodesics.
To see that the manifold does not end there, the authors use
the modified radial coordinate
${r'}^2 = r - m$, and show that the geometry
contains an event horizon at $r'=0$ but has no singularity
(Fig. 2).  It is interesting to note that this is identical to the
causal structure seen in the extremal Kerr solution along the axis
of symmetry \cite{H&E}.

Finally, for the case $|Q|>m$ the authors find that the
manifold is completely regular, when using the appropriate
radial coordinate ${\tilde r}^2=r-Q^2/m$. It terminates at $\tilde r=0$,
and the potential conical singularity there is removed by a periodic
identification of the spacelike coordinate $x$. Therefore, the spacelike
sections have the structure of a ``cigar'', looking flat near the origin
but asymptotically approaching ${\bf R} \times {\bf S}^1$.

Let us now turn our attention to our new solutions.
Both (\ref{w10}) and (\ref{w17}) share some
of the features of the static black string (\ref{w7}).
They are both asymptotically flat configurations with
two Killing fields, $\di_x$ and $\di_t$, with infinity described by
the limit $\tilde r = \ln \rh \rightarrow \infty$,
where they approach exponentially fast the linear dilaton vacuum,
with flat Minkowski metric and vanishing gauge fields.
They also have three metric
singularities each, $\rh=m\ch^2$, $m\sh^2 + Q^2/m$, and $0$
for the first, and  $\rh=\mu b^2 + q^2/\mu$, $b^2 q^2/\mu + \mu$,
and again $0$ for the second.
The surface $\rh = m\sh^2$ in the first solution actually isn't singular, as
can be seen from expanding the squared bracket in (\ref{w10}) and
collecting the like terms. The nature of the singular points can be examined
by investigating the behavior of curvature invariants as these
points are approached. This arduous task is in fact easier in
three dimensions, because Weyl curvature is identically zero,
and the only scalar curvature invariants are $R$, $R^{\mu\nu}R_{\mu\nu}$
and $\det(R_{\mu\nu})/\det(g_{\mu\nu})$ \cite{Wein}, which all
blow up as $\rh \rightarrow 0$ and are finite elsewhere.
Thus $\rh=0$ is the
only polynomial curvature singularity for our solutions.
We should note here that our choice to rely on the
conventional definition of curvature singularities of General Relativity
is equivalent to assuming that the space-time geometry can be probed
only by pointlike observers. Whereas this assumption is obviously of limited
validity in string theory, it is a useful working tool
in the absence of a more general definition, and we will restrict our
attention to it (for more general criticism see Ref. \cite{HorTse}).

In order to analyze our solutions further, we have to investigate them
one by one. The results are summarized in the following five subsections.

\subsection{The first family with $0<|Q| < m$}

Here we present our first black string solution.
The surfaces $\rh_{+} = m\ch^2$ and
$\rh_{-} = m\sh^2 + Q^2/m$ are removable singularities,
where coordinates change signature, and thus represent the event and Cauchy
horizons, respectively. This can be seen from the fact that they are
both null surfaces, and that almost
all timelike and null geodesics cross them,
as we will demonstrate shortly. The behaviour of the coordinates
while crossing these surfaces is somewhat different from the situation
enjoyed by the Horne-Horowitz black string. While the radial coordinate
behaves the same, being spacelike outside the event horizon and inside
the Cauchy horizon, and timelike in between, the time at infinity $t$,
which turns spacelike after crossing the event horizon, regains the timelike
character again after crossing the interior static limit $\rh_{t} = m\sh^2$,
inside the Cauchy horizon. Likewise, $x$ also changes signature,
becoming timelike after the surface
$\rh_{x} = Q^2/2m + (Q^2 \sh^2\ch^2 + Q^4/4m^2)^{1/2}$.
Thus, as in the static black string case, representation of
the causal structure by planar Penrose diagrams is not completely
accurate, since there is more freedom in choosing the time coordinate,
but the situation here is a bit more complicated.
As we see, there are regions where the time coordinate is
an $r$-dependent linear combination of $t$ and $x$.
However, if we keep this in mind, we can still employ the
diagrammatic technique as a descriptive tool.

Before completing the description of the causal structure, we
will investigate geodesics of this solution. Again, due to the presence
of two Killing vector fields, the geodesic equations take a particularly
simple form. Introducing two integrals of motion associated with
the cyclic coordinates $P_{\mu} = (-E,P)$, and the squared rest mass of the
particle moving on the geodesic $p = 0,1$ (distinguishing null and timelike
geodesics),
we obtain the following
formula for the radial coordinate (the overdot denotes the derivative
with respect to the affine parameter):
\bea \label{w18}
\frac{{\dot{\rh}}^2}{2\Lambda \rh^2}&=& E^2
\bigl(1-\frac{Q^2}{m\rh}-\frac{Q^2\sh^2\ch^2}{\rh^2}\bigr)
- P^2\bigl(1-\frac{m\ch^2}{\rh}\bigr)
\bigl(1-\frac{m\sh^2}{\rh}\bigr)  \nonumber \\
&&~~~~~~~~~~~~ + 2EP \frac{Q\sh^2}{\rh}\bigl(1-\frac{m\ch^2}{\rh}\bigr)
- p\bigl(1-\frac{m\ch^2}{\rh}\bigr)
\bigl(1-\frac{m\sh^2+Q^2/m}{\rh}\bigr).
\eea
Upon the inspection of this equation, we note
that while all causal geodesics cross the event
horizon, the subset for which $m\sh^2 E = QP$
terminates at the Cauchy horizon. All other causal geodesics pass through
the Cauchy horizon too, and specifically null geodesics with $P=0$
terminate at the
surface $\rh_{x} < \rh_{-}$, where $\partial_x$ becomes null,
(but does not vanish, as can be seen from computing the component of
the tangent along $x$ there). We note that the behavior
of the $m\sh^2 E = QP$ geodesics is related to the $P=0$ case studied
in the static solution by Horne and Horowitz. They found that $P=0$ geodesics
terminate at the Cauchy horizon, and ascribed this to the fact that
there is a world line along which the field $\partial_x$ must be identically
zero, and not just null. Note that
for our solution, the $P=0$ null geodesics are protected from this
by the terms proportional to $\sh^2 > 0$, but that we recover the pathology
in the limit $\sh^2 \rightarrow 0$, when the above two geodesics coincide.
Therefore we see that the cross-term in our metric
has caused the pathological class of geodesics to shift from $P=0$
to $m\sh^2 E = QP$. They end at the equivalent region of the black string,
with the only difference that now it is the vector
$Q \partial_x + m \sh^2 \partial_t$, that vanishes there, because
it is orthogonal everywhere to the class $m\sh^2E=QP$ but becomes null
on the Cauchy horizon and timelike inside of it.

In order to see what happens in the region near the singularity, it is
helpful to rewrite the radial geodesic equation (\ref{w18}) by collecting
the terms of the same order of divergence:
\bea \label{w19}
\frac{{\dot{\rh}}^2}{2\Lambda \rh^2}&=& E^2 - P^2 - p -
\frac{(QE + mP)(QE - m(\ch^2 + \sh^2)P) - pm(\ch^2 + \sh^2)}
{m\rh}~~~~~~~~~ \nonumber \\
&&~~~~~~~~~~~~~~~~~~~~~~~~~~~~~~~~~~
- \frac{(QE + mP)^2\sh^2\ch^2 + pm\ch^2(m\sh^2 + Q^2/m)}{\rh^2}.
\eea
The coefficient of the $O(1/\rh^2)$ term in this equation
is nonpositive for all causal geodesics. As a consequence,
no causal geodesic with this term being nonzero, beginning outside of the
black string, can reach the singularity at $\rh = 0$, because
the $O(1/\rh^2)$ term forces it to stop and turn. Thus, the only
geodesics which don't turn away from the singularity are
null geodesics with $QE + mP = 0$.
This is somewhat reminiscent of the behavior of geodesics in the Kerr
solution. Noting that $Q$ is similar to angular momentum in this geometry,
we can define $a=Q/m$ in analogy with the angular momentum parameter
in the Kerr solution. For the geodesics in the equatorial plane
of the Kerr solution we can define
the impact parameter $b=l/E$, where $l$ is the conserved angular parameter
analogous to our $P$. For the value of
the impact parameter $l/E=a$ these geodesics hit the ring singularity,
and in essence behave in the same way as all radial geodesics do in
static black hole spacetimes. Thus, we see that our condition
$QE + mP = 0$ is analogous to the Kerr case, again singling out only
those radial geodesics which reach the singularity.

There is a startling difference between our case and Kerr, however.
In Kerr, geodesics with $b=a$ are linear in the affine parameter,
$r=E \lambda + {\rm const}$. In contrast, in our case the equation
(\ref{w18}) reduces to the
standard linear homogenous equation, with exponential solutions
$\rh \propto \exp(\lambda)$, (analogous to the case $|P/E|=|Q/M|$
for the static solution (\ref{w7})).
Thus, although the singularity is the attractor for these geodesics,
as they can come arbitrarily close to it, they cannot reach it
for any finite value of the affine parameter. As a consequence, our spacetime
is timelike {\it and} null geodesically complete. The singularity still
must be included in the manifold, which is
spacelike geodesically incomplete. In addition, it can also be
reached by nongeodesic causal curves. Yet, it is
quite harmless for pointlike observers, living serenely along causal
geodesics.  Remarkably, it would appear to an observer
inside the Cauchy horizon as some eerie but ultimate
warning against dangerous living!

Calculation of the Hawking temperature for this solution is complicated
by the presence of cross-terms in the metric. Employing the approach of
\cite{YBM}, designed for such situations, we can obtain it by rewriting
the metric (\ref{w10}) in the ADM form, and then Wick-rotating the time
coordinate $t=i\vartheta$. Requiring that the horizon is a regular point,
we must identify $\vartheta$ with the period
$2\sqrt{2}\pi m\ch^2/\sqrt{\Lambda(m^2-Q^2)}$. This gives the following
expression for the Hawking temperature:
\be \label{Hawk1}
T= \sqrt{\frac{\Lambda}{2}} \frac{\sqrt{m^2-Q^2}}{2\pi m \ch^2}.
\ee
As in the static black string and the Reissner-Nordstr$\o$m case, as
$Q^2 \rightarrow m^2$, the temperature vanishes. Thus the string would
settle down to $|Q|=m$ in the absence of charge-dissipating processes.

Finally, as we have indicated above, the solution has a static limit
at $\rh_{t}$, where the coordinate $t$ again becomes timelike. Thus,
inside this surface it is again possible to find observers at rest with
respect to the asymptotic infinity, much like the Reissner-Nordstr$\o$m
solution, and unlike the static black string of Horne and Horowitz.
In conclusion, the causal structure of this solution up to the Cauchy horizon,
is qualitatively similar to that of the Horne-Horowitz black string, with
the differences arising near the singularity. The corresponding diagram is
presented in Fig. 1.

\subsection{The extremal limit $|Q| = m$ of the first family}

As usual, we define the extremal limit of our black string by a
choice of parameters which ensures the coincidence of the
two horizons $\rh_{-} = \rh_{+}$. Naively, we would then expect to
obtain a solution with a singularity enclosed by a single horizon.
However, Horne and Horowitz found that in the corresponding static case
(\ref{w7}), the coordinate $\rh$ was not the proper
extension across the horizon $\rh=m$. A hint that a different extension
was needed was provided by the radial geodesic equation, which indicated
that the horizon is a radial turning point for all causal geodesics.
In analogy with this situation,
we find that (\ref{w10}) does not give the correct
extension across the horizon $\rh=m\ch^2$ in the extremal case.
Namely, the radial geodesic equation (\ref{w18}) for the extremal case
$|Q|=m$ can be rewritten as
\be \label{w20}
\frac{\dot{\rh}^2}{2\Lambda} = (\rh-m\ch^2)\Bigl((E+P)
\bigl((E-P)\rh + m\sh^2(E+P)\bigr) - p(\rh - m\ch^2)\Bigr).
\ee
The right hand side of
this equation vanishes at the horizon, and thus it appears that
no timelike or null geodesics can cross
this horizon. To rectify this problem, we follow the approach of \cite{HorHo}
and define the new radial coordinate $\rb^2=\rh-m\ch^2$.
In terms of it, the radial equation becomes
\be \label{w21}
\frac{2\dot{\rb}^2}{\Lambda} = (E+P)
\bigl((E-P)(\rb^2 + m\ch^2) + m\sh^2(E+P)\bigr) - p\rb^2,
\ee
and thus we see that all causal geodesics ($p \ge 0$) in fact cross the
horizon,
located at $\rb =0$.
In terms of this coordinate the metric takes the form
\be \label{w21a}
    ds^2=\frac{2}{\Lambda}\frac{d\rb^2}{\rb^2} + \frac{\rb^2}{\rb^2+m} dx^2
        -\frac{\rb^2}{(\rb^2+m)(\rb^2+m\ch^2)^2}\,((\rb^2+m)dt-m\sh^2dx)^2.
\ee

As in the static solution (\ref{w7}), the only metric singularity is
at the horizon $\rb=0$. This, of course, is a removable singularity,
and the metric can be extended beyond it, to the region
$\rb <0$. Furthermore, (\ref{w21a}) is
invariant under the reflection $\rb \rightarrow - \rb$. Thus, after
passing through $\rb = 0$ a particle finds itself in a
universe identical to the one which it just left. As a
consequence, the maximal extension of this solution is based on
a zig-zag event horizon along which an infinite number of asymptotically
flat regions are connected, with the causal structure identical to the
extremal black string of Horne and Horowitz (Fig. 2).

\subsection{The first family with $|Q| >m$}

In this case, the signature of the metric changes at the
surface $\rh=m\sh^2 + Q^2/m$, since the change of sign of the metric component
$g_{\rh\rh}$ is accompanied by the change of both eigenvalues of
$g_{2}$ to negative values, as can be seen from (\ref{w10}).
The overall signature change is from (-,+,+) to (-,-,-).
This indicates that the metric (\ref{w10})
cannot be extended beyond $\rh=m\sh^2 + Q^2/m$.

To obtain the correct picture,
we must redefine the radial coordinate.
This time, we employ $\rb^2=\rh - m\sh^2 - Q^2/m$. With this,
our metric becomes
\be  \label{w22}
    ds^2=\frac{2}{\Lambda}\frac{d\rb^2}{\rb^2+\frac{Q^2}{m}-m} +
\frac{\rb^2}{\rb^2+\frac{Q^2}{m}} dx^2
-\frac{\rb^2+\frac{Q^2}{m}-m}{(\rb^2+\frac{Q^2}{m})
(\rb^2+\frac{Q^2}{m}+m\sh^2)^2}
        \,((\rb^2+\frac{Q^2}{m})dt - Q\sh^2 dx)^2.
\ee
In these coordinates, the surface $\rb = 0$ is singular, since the
metric is degenerate there. In fact, if we expand this metric near the
origin, after introducing a new radial coordinate by
$\rb = \sqrt{(Q^2/m-m)}\sinh(\sqrt{\Lambda/2} z)$, we obtain
\bea  \label{w23}
    ds^2&=&dz^2 + \frac{\Lambda}{2}\frac{Q^2-m^2}{Q^2}z^2dx^2 \nonumber \\
&-& \frac{Q^2(Q^2-m^2)}{(Q^2+m^2\sh^2)^2}
\Bigl(1+\Lambda z^2(\frac{m^2\ch^4}{Q^2 + m^2\sh^2}
-\frac{m^2}{2Q^2})\Bigr)
\Bigl(dt - \frac{m\sh^2}{Q}
(1-\frac{\Lambda}{2}\frac{Q^2-m^2}{Q^2}z^2)dx\Bigr)^2 \\
&+& O(z^4),\nonumber
\eea
as $z \rightarrow 0$.
This metric looks like the metric of a spinning point source in
three dimensions \cite{DJH},
and hence $\rb=0$ represents the standard coordinate
singularity at the origin, provided that we have smoothed it
by identifying $x$ with the same period
$\Pi_x = 2\sqrt{2}Q \pi /\sqrt{\Lambda(Q^2 - m^2)}$ as discussed in
\cite{HorHo}.
Otherwise, we would have ended up with a conical singularity there.

The global structure of this manifold is considerably different
from the static case. The manifold can still be thought of
as consisting of infinite ``cigar''-shaped spatial hypersurfaces
defined by adjusting the time coordinate such
that $(\rb^2 + Q^2/m)dt - Q\sh^2 dx=0$, (or generated
by spacelike geodesics $\dot t = \dot x = 0$), planar near the
origin and deforming towards ${\bf R}\times {\bf S}^1$ as
$\rb \rightarrow \infty$. However, in this solution  the angular
coordinate $x$ is ``bolted'' to time in a nontrivial manner, and hence
there now appear closed timelike curves.
This can be seen by realizing that the coordinate $x$
becomes null at the surface
${\rb_x}^2 = \sqrt{Q^4/4m^2 + Q^2 \sh^2\ch^2} - Q^2/2m - m\sh^2 > 0$, and thus
for $\rb < \rb_x$ the loops $(\rb = {\rm const.}$,
$t={\rm const.})$ are timelike. Furthermore, there
are geodesics which can reach this region.
The geodesic equations in this case can
be rewritten in a particularly convenient form by introducing local
coordinates $\tilde x = x - P\lambda$ and $\tilde t = t - m\sh^2 x/Q$.
These coordinates span a helical frame along each geodesic, twisting
around it as the affine parameter $\lambda$ changes. Then,
using $L=Q(QP-m\sh^2E)/m$, we get
\newpage
\bea \label{w24}
\dot{\tilde x} &=& \frac{L}{\rb^2}, \nonumber \\
\dot{\tilde t} &=& - \frac{m\ch^4}{\rb^2 + Q^2/m - m} + E, \\
\frac{2}{\Lambda} {\dot{\rb}}^2 &=& (\rb^2 + Q^2/m - m)(E^2 - P^2 -p)
+ \frac{m}{Q^2}(Q^2\ch^4E^2 -L^2) - \frac{Q^2 - m^2}{Q^2}\frac{L^2}{\rb^2}.
\nonumber
\eea
If we ignore the first terms in the $\tilde t$ and $\rb$
equations, we obtain precisely the polar parametrization of straight lines
in Minkowski space. The parameter $L$ then represents the conserved
angular momentum along the lines, preventing them from hitting the origin
unless $L=0$. Thus we see that the curvature effects are described
by the first terms in the last two equations of (\ref{w24}), and that
their effects (other than rescaling the constant parameters) are essentially
negligible near the origin, as indicated by the expansion (\ref{w23}).
Moreover, comparing the $L$-dependent terms we confirm our choice
of compactification of the coordinate $x$.

To shed more light on this geometry we can look at several typical
geodesics. The first natural candidate is,
of course, the $L=0$ case, generalizing
rays through the origin from the flat background. We note that in terms
of the integrals of motion $E$ and $P$ this condition translates to
$QP=m\sh^2E$. In the static case, when $\sh^2=0$, these are just the
lines of constant $x$ and of infinite span in $\rb$, which pass through
the origin and escape to infinity on both sides. In our case, when
$\sh^2 \ne 0$, this picture is correct only for $\sh^2 \le |Q|/m$; if
reparametrized in terms of the original coordinate $x$, these trajectories
are hyperbolic spirals, approaching the spiral of Archimedes as
$\sh^2 \rightarrow |Q|/m$. The main point is that these geodesics enter
and exit the region of space-time with timelike loops.

In contrast, when $\sh^2 > |Q|/m$, we have
$E^2-P^2-p=E^2(1-m^2\sh^4/Q^2)-p<0$, and all causal geodesics of this
kind are bound orbits oscillating near the origin, looking
like $\rb \propto \sin(x)$ in the original variables.
Their amplitude is bounded from above by
${\rb_{max}}^2 = (m^2\sh^2+Q^2)/m(m^2\sh^4-Q^2)$, and for large enough
$\sh^2$ they remain within the region with closed timelike curves.
Nonetheless, communication between the two
regions is still possible. For example,
null geodesics with $P=E$, which correspond to straight lines with
impact parameter $l^2$ in the flat space, can reach into the region
with timelike loops. Their closest approach to the origin is
given by the minimal impact parameter
${l_{min}}^2=(Q+m)(Q+m\sh^2)^2/m\sh^2(2Q + (Q+m)\sh^2)$,
which is less than ${\rb_x}^2$, as can be seen from $g_{xx}(l_{min})<0$.

Thus we conclude that the two regions are always geodesically connected, and
the region with closed timelike curves cannot be smoothly detached
away from the manifold. Because we can extend this solution
to four dimensions, by simply adding an additional flat coordinate, it
might be interesting as an example of a spacetime which allows
time travel. However, its actual physical significance would remain
somewhat dubious, due to its asymptotic topology.

\subsection{Second family of black strings}
Before proceeding with the analysis,
it is useful to rewrite the solution (\ref{w17}) using
a different set of parameters.
Specifically, we eliminate the parameters $\mu$, $q$ and $b$ in favor
of the world-sheet frame ADM mass and linear momentum in the $x$ direction, as
well as the electric and axionic charges, defined by Gauss laws
for the two fields. The ADM parameters are the components of the flux of
linearized energy-momentum tensor integrated over a spacelike hypersurface
at infinity, where the metric can be expanded around the Minkowski form:
$g_{\mu\nu} = \eta_{\mu\nu} + \gamma_{\mu\nu}$.
The necessary formulae are
given in \cite{HHS}, which in our case give the following expressions
for these quantities per unit length $x$ of the string (with the Gauss
laws, which are the integrals of conserved currents over the same
spacelike hypersurfaces, given here in form notation),
\bea \label{charges}
{\cal M} &=& - \frac{1}{2\Lambda}e^{-\Phi}(\gamma_{xx}^{'}
+ \gamma_{rr} \Phi'), ~~~~~~~~~~
{\cal P}_{x} = - \frac{1}{2\Lambda}e^{-\Phi}\gamma_{tx}^{'}, \nonumber \\
e &=& \frac{1}{2\sqrt{2}\Lambda}e^{-\Phi}{^{*}}{\bf F},
{}~~~~~~~~~~~~~~~~~~~~~~ {\cal Q} = \frac{1}{2\Lambda}e^{-\Phi} {^{*}}{\bf H} .
\eea
The prime denotes the derivative with respect to the
``flat'' radial coordinate $z=\ln(\rh)/\sqrt{2\Lambda}$, and the
additional $\sqrt{2}$ in the definition of the electric charge $e$
reflects our normalization of the $F^2$ term in the action (\ref{w1}).
This gives ${\cal M} = Mb^2$, ${\cal P}_x = 0$, $e = qb\sqrt{b^2 -1}$
and ${\cal Q} = bM$. Now we can solve for $\mu$, $q$ and $b$, to obtain
$\mu = {\cal Q} (\sqrt{{\cal M}^2 - {\cal Q}^2}\pm
\sqrt{{\cal M}^2 - {\cal Q}^2 - 4e^2})
/2{\cal M}\sqrt{{\cal M}^2 - {\cal Q}^2}$,
$q = {\cal Q}^2 e/{\cal M}\sqrt{{\cal M}^2 - {\cal Q}^2}$,
$b = {\cal M}/{\cal Q}$, and finally $M = {\cal Q}^2/{\cal M}$.
Using these parameters, we can rewrite our second family of
solutions (\ref{w17}) as
\bea \label{w17adm}
ds^2 &=&
\frac{d\rh^2}{2\Lambda(\rh-\rh_+)(\rh-\rh_-)} +
(1-\frac{{\cal Q}^2}{{\cal M}\rh}
-\frac{e^2{\cal Q}^2}{{\cal M}^2\rh^2}) dx^2 ~~~~~~~ \nonumber \\
&&~~~~~~~~~~~~~~~~~~~~~~~~- (1-\frac{{\cal M}}{\rh}
+\frac{e^2}{\rh^2})dt^2 -
     2 \frac{{\cal Q} e^2}{{\cal M} \rh^2} dxdt, \nonumber \\
   &&~ B = \frac{{\cal Q}}{\rh} dx \wedge dt, ~~~~~~~~~~~~~
A = - \sqrt{2}\frac{e}{\rh}(dt + \frac{\cal Q}{\cal M} dx),  \\
 &&~~~~~~~~~~~~~~~~~~~~~~~  e^{-\Phi}= \sqrt{2 \Lambda} \rh, \nonumber
\eea
with the horizons given by
$\rh_{\pm} = ({\cal M}^2 + {\cal Q}^2 \pm \sqrt{{\cal M}^2 - {\cal Q}^2}
\sqrt{{\cal M}^2 - {\cal Q}^2 - 4e^2})/2{\cal M}$. We note the distinct
appearance of the factor of $4$ together with $e^2$ here. This is, as
we have pointed out above, due to
our normalization conventions for the gauge field ${\bf F}$. We should also
point out that similar variables for our first family of solutions give
a representation far less transparent than the one provided by (\ref{w10}).
This comes about because the first family
has nonvanishing momentum ${\cal P}_x$, which is non-trivially related
to the axion charge \cite{HHS}.

 From the formula for $\rh_{\pm}$
we can now determine the range of parameters which split this
family into different subclasses. Obviously the possibilities are
${\cal M}^2 \ge {\cal Q}^2 + 4e^2$,
${\cal Q}^2 <{\cal M}^2 < {\cal Q}^2 + 4e^2$ and
${\cal M}^2 \le {\cal Q}^2$.
In the last two cases, although the square roots which
appear in the definition of $\rh_{\pm}$ become imaginary,
the denominator of the lapse function (which is the only part of the solution
containing explicit reference to these terms) remains real, as can be
readily verified. Thus, in general, these two cases cannot be excluded.
We will not study their properties in detail for the following reasons.
In the case defined by the second inequality,
${\cal Q}^2 <{\cal M}^2 < {\cal Q}^2 + 4e^2$, we note that both $\rh_{\pm}$
are complex numbers. Thus the metric is regular everywhere except at $\rh=0$,
where we have found a curvature singularity. As a consequence, this
solution describes a geometry containing a naked singularity, much like
the Reissner-Nordstr$\o$m solution with $M^2 < e^2$.
Furthermore, because $b={\cal M}/{\cal Q}$, and
recalling that the $b<1$ case is related to $b>1$ by a simultaneous Wick
rotation of both $t,x$ variables, we see that the case represented
by the last inequality is the proper extension of the solution to
$b<1$. Drawing on the similar relationship between
the first and the third subclass of
our first family of solutions, we conclude that this case must
be similar to the $|Q|>m$ subclass of our first family of solutions,
containing closed timelike curves,
wherefore we will not elaborate it further.

In the remainder of this section,
we will concentrate on the first inequality as well as the two equalities,
${\cal M}^2 = {\cal Q}^2 + 4e^2$, and ${\cal M}^2 = {\cal Q}^2$.
We will first elaborate the properties of the non-extremal subclass
${\cal M}^2 > {\cal Q}^2 + 4e^2$. Here we find a surprisingly rich geometric
structure, which looks like a hybrid of black holes
in four dimensions and our first family of solutions.

To start with, we observe that this solution again possesses the event
horizon and the Cauchy horizon, $\rh_{+}>\rh_{-}$ respectively. All causal
geodesics starting from infinity cross $\rh_{+}$, while there still exists
the pathological class of geodesics which terminates at the Cauchy horizon,
much like the previously discussed cases. This can be seen as follows.
After the integrals of motion $P_{\mu}=(-E,P)$ and the squared rest mass of the
particle $p$ are introduced, the radial geodesic equation can be written as
(again the overdot denotes the derivative with respect to the
affine parameter):
\bea \label{w25}
\frac{{\dot{\rh}}^2}{2\Lambda \rh^2}&=& E^2 - P^2 - p -
\frac{({\cal Q}E + {\cal M}P)({\cal Q}E - {\cal M}P)
- p({\cal M}^2+{\cal Q}^2)/{\cal M}}{{\cal M}\rh}~~~~~~~~~ \nonumber \\
&&~~~~~~~~~~~~~~~~~~~~~~~~~~~~~~~~~~
- \frac{e^2({\cal Q}E + {\cal M}P)^2
+ p({\cal M}^2{\cal Q}^2 + e^2({\cal M}^2 + {\cal Q}^2))}{{\cal M}^2\rh^2}.
\eea
The terms proportional to the squared rest mass of the probe $p$
do not affect the properties of geodesics near the two horizons. Ignoring
them, we can employ the radial coordinate shifted by the value of the event
horizon $\rho = \rh - \rh_{+}$. We can then rewrite this equation after
introducing the parameters
$p_{\pm} = (\sqrt{{\cal M}^2 - {\cal Q}^2}\pm
\sqrt{{\cal M}^2-{\cal Q}^2-4e^2})/2$ as
\bea \label{w26}
\frac{{\dot{\rho}}^2}{2\Lambda}&=&\frac{(p_{+}{\cal M}E-p_{-}{\cal Q}P)^2}
{{\cal M}^2} + (E^2-P^2)\rho^2 \nonumber \\
&&~~~~~~~~~~~~~~~~+ \frac{{\cal M}^2E^2-{\cal Q}^2P^2}
{{\cal M}}\rho + \frac{(E^2-P^2)\sqrt{{\cal M}^2-{\cal Q}^2}
\sqrt{{\cal M}^2-{\cal Q}^2-4e^2}}{{\cal M}}\rho.
\eea
For all inwards-oriented geodesics which emanate
from infinity ($E^2\ge P^2$) the
RHS of this equation never vanishes for any $\rho \ge 0$.
Thus they all cross the event horizon and fall into the black string.

Similarly, we can use the radial coordinate shifted at
the Cauchy horizon, defining
$\tilde \rho = \rh - \rh_{-}$. The radial equation becomes
\bea \label{w27}
\frac{{\dot{\tilde \rho}}^2}{2\Lambda}&=&
\frac{(p_{-}{\cal M}E-p_{+}{\cal Q}P)^2}
{{\cal M}^2} + (E^2-P^2){\tilde \rho}^2 \nonumber \\
&&~~~~~~~~~~~~~~~~+ \frac{{\cal M}^2E^2-{\cal Q}^2P^2}
{{\cal M}}\tilde \rho - \frac{(E^2-P^2)\sqrt{{\cal M}^2-{\cal Q}^2}
\sqrt{{\cal M}^2-{\cal Q}^2-4e^2}}{{\cal M}}\tilde \rho.
\eea
Again, we look only at the arcs of geodesics outside of the Cauchy horizon;
hence $\tilde \rho \ge 0$. Because the coefficient of the linear term in
$\tilde \rho$ is always positive, the RHS of this equation vanishes only
when $\tilde \rho =0$ and $p_{-}{\cal M}E=p_{+}{\cal Q}P$ simultaneously.
The last condition is compatible with $E^2 \ge P^2$ since $p_{-}<p_{+}$.
Therefore, we see that the class of geodesics for which
$p_{-}{\cal M}E=p_{+}{\cal Q}P$ stops at the Cauchy horizon. This corresponds
exactly to the case $m\sh^2E=QP$ studied in the first family of black strings,
and shows that the pathology found by Horne and Horowitz still persists.

Another similarity between this solution and our first family is that this
black string is also causally geodesically complete. Once again, the only
causal geodesics which approach the singularity without turning
are null geodesics for which
${\cal Q}E + {\cal M}P = 0$, which come arbitrarily close to the
singularity but again according to $\rh \propto \exp{\lambda}$.
All other causal geodesics turn at a finite $\rh >0$, where the
repulsive term of order $O(1/\rh^2)$ in (\ref{w25}) prevails.
Thus, the singularity can never be reached by any causal geodesics for
finite value of the affine parameter, and it appears very much the same as
in the first family of black strings.

There are, however, considerable differences between the two families.
Namely, the second family (\ref{w17adm}) possesses three Killing horizons,
$\rh_{E\pm}=({\cal M} \pm \sqrt{{\cal M}^2-4e^2})/2$ and
$\rh_{x}={\cal Q}({\cal Q} + \sqrt{{\cal Q}^2 + 4e^2})/2{\cal M}$
where the metric is regular but one of the
coordinates $t,x$ becomes null. The first two, where $t$ is null,
satisfy $\rh_{E+} > \rh_{+} > \rh_{-} > \rh_{E-}$,
and resemble the situation found in the Kerr black hole in four dimensions.
The location of the last Killing horizon, where $x$ is null,
is inside the event horizon, but depending on the values of ${\cal M}$,
${\cal Q}$ and $e$ it can be either inside or outside the Cauchy horizon.

The outer Killing horizon $\rh_{E+}$ defines the ergosphere, and thus one might
expect that there exist Penrose-type processes for energy extraction
from this kind of black string. This issue is far from clear-cut, though,
because the energy extracted from the Kerr black hole is at the expense of
the hole's momentum, resulting in slow-down of its rotation, and
disappearance of the ergosphere.
In our case, quite unexpectedly,
the ergosphere appears due to the charges of the axion and
gauge fields, which are protected by the Gauss laws at infinity. Therefore,
energy extraction by a Penrose-type process would seem to be inextricably
linked to the diminishing of the string's charge, which is in
contradiction with the Gauss laws. We believe that the consistent
resolution of this problem should be sought by postulating the
existence of a more general family of solutions, which
will be characterized by a non-zero linear momentum along the string
${\cal P}_{x}$. This quantity would then be dissipated by Penrose
processes, thus opening the channel for eliminating the ergosphere
while keeping the gauge charges conserved.
A more detailed study of this problem would seem to be merited.
Ultimately, one would like to determine the set of all alowed conserved
quantities for three-dimensional stationary configurations, analogous
to the approach of the last of Ref. \cite{CC}.

Finally we present the Hawking temperature for this solution. It is obtained
analogously to (\ref{Hawk1}), and is given by
\be \label{Hawk2}
T = \sqrt{\frac{\Lambda}{2}}
\frac{\sqrt{{\cal M}^2-{\cal Q}^2}\sqrt{{\cal M}^2-{\cal Q}^2-4e^2}}
{\sqrt{{\cal M}^2-{\cal Q}^2} + \sqrt{{\cal M}^2-{\cal Q}^2-4e^2}}.
\ee
In this case, the Hawking temperature vanishes for both extremal limits
${\cal M}^2 = {\cal Q}^2 + 4e^2$ and ${\cal Q}^2 = {\cal M}^2$.
Which of these limiting situations will be reached by evaporation
could in principle be determined by a detailed study of linear momentum
transfer between the black string and the Hawking radiation, which is
beyond the scope of the present work.

In sum, the causal structure of this solution is reminiscent of the
first family. The associated Penrose diagram is essentially the same.
The most important difference is the appearance of the
ergosphere, which can provide for interesting effects in this geometry.
The causal structure for this case is also shown in Fig. 1.

\subsection{Extremal limit(s) of the second family}
As we have indicated above, we will look here at the two special cases of the
second family of solutions. We refer to these as the extremal limits in a
somewhat tentative manner, because they represent such choices of parameters
where the two horizons become degenerate. Yet, the case
${\cal Q}^2 = {\cal M}^2$ deserves its label as an extremal black string
only in an indirect fashion, as we will indicate below, and show in the next
section.

Our first extremal limit is given by the condition
${\cal M}^2 = {\cal Q}^2 + 4e^2$, resembling the extremality condition for
dyonic Reissner-Nordstr$\o$m black holes.
This case is very different from the previously studied extremal limits.
There is now a singularity at $\rh=0$, a single degenerate horizon
$\rh_{h}=({\cal M}^2 + {\cal Q}^2)/{\cal M}$ and, in general,
three Killing horizons, located at
$\rh_{E\pm}=({\cal M}\pm{\cal Q})/2$ and
$\rh_{x}=({\cal M}{\cal Q}+{\cal Q}^2)/2{\cal M}$. By comparing the values
of the parameters, we see that there is an ergosphere $\rh_{E+}$ outside
of the event horizon $\rh_{h}$, and that the remaining two Killing horizons
are inside of $\rh_{h}$. Their relative locations however depend on the ratio
${\cal Q}/{\cal M} <1$, and they coincide for
${\cal Q}/{\cal M} = \sqrt{2}-1$. To see that all of these are indeed
contained in the manifold, in contrast to our first extremal limit
and the extremal limit in the static case,
we need to look at the geodesic equations and demonstrate that
there are geodesics which extend to all of the above surfaces.
This is again controlled by the radial equation.
Qualitatively the behavior of geodesics is the same as in the non-extremal
case. Here we will just show that all geodesics cross the event horizon in
original coordinate $\rh$, meaning that the proper extension is given
by including in the manifold the sector with $\rh<\rh_{h}$.
Since in the extremal
limit, the parameters $p_{\pm}$ degenerate to $e$, we can rewrite the
radial equation for null geodesics, in terms
of the radial coordinate shifted by the horizon $\rho=\rh-\rh_{h}$, as
\be \label{w28}
\frac{{\dot{\rho}}^2}{2\Lambda}=\frac{e^2}{{\cal M}^2}
({\cal M}E-{\cal Q}P)^2 + (E^2-P^2)\rho^2
+ \frac{{\cal M}^2E^2-{\cal Q}^2P^2}
{{\cal M}}\rho.
\ee
The RHS of this equation does not vanish for any $\rho\ge 0$ (i.e., outside
of the event horizon), and since similar conclusion also holds for
timelike geodesics, we see that all inwards-oriented geodesics fall into
the black string, proceeding to $\rho<0$, as claimed.

To see that the other characteristic surfaces are also
reachable, we need only observe that there still exists the class
of null geodesics with ${\cal Q}E+{\cal M}P=0$, discussed in the non-extremal
black string background. These are the only causal geodesics in the
manifold that come arbitrarily close to the singularity
at $\rh=0$. However, since their descent towards the singularity is controlled
by the exponential of the affine parameter, they do not reach it in any
finite range of the parameter, and thus this manifold is also causally
geodesically complete. Thus we conclude that the causal structure of
this geometry is similar to the extremal Reissner-Nordstr$\o$m solution,
the differences being the ergosphere and the causal geodesic
completeness of our solution. This comparison of causal structure is
graphically summarized by the Penrose diagram in Fig. 3.

We should also point out here that this solution does not have any null
Killing vectors, in contrast to other extremal black strings. This can be
seen from noting that any Killing vector must be a linear combination
$\alpha \partial_{x}+\beta \partial_{t}$. The null condition for this
solution then translates into $\alpha^2=\beta^2$ and
$\alpha{\cal Q} + \beta{\cal M} =0$. These two
equations can be simultaneously solved only if ${\cal Q} = \pm {\cal M}$,
which is not the case here.

The other extremal limit, ${\cal Q}^2={\cal M}^2$, as we have mentioned
above, can be interpreted as a black string only indirectly. For example, we
can see from the geodesic equation for the radial coordinate
(\ref{w28}) for this case, that no causal geodesics starting from infinity
can ever cross the horizon, although some may come arbitrarily close to
it before bouncing back. Nevertheless, if the horizon were
probed by non-geodesics world lines, one would discover a structure akin
to the horizon of the extremal static case or the extremal limit of
our first family of solutions, where the geometry consists of two mirror
images divided by the horizon (Fig. 2). In effect, in this case the geodesics
encounter an infinite potential barrier at the boundary, due to the terms
in the metric proportional to the electric charge. We will present these
arguments in mathematical form in the next section, in a slightly more
general context.

The analogy of this case
with an extremal black string can be further strengthened if we employ the
null coordinates $u=(t\pm x)/\sqrt{2}$, $v=(t\mp x)/\sqrt{2}$, where the
signs are chosen according to whether ${\cal Q}/{\cal M}$ is positive or
negative unity, respectively. With these coordinates, we can rewrite the
solution as
\bea \label{29}
ds^2 &=&
\frac{d\rh^2}{2\Lambda(\rh-{\cal M})^2} -2(1-\frac{{\cal M}}{\rh})dudv
-\frac{e^2}{\rh^2}du^2, \nonumber \\
   &&~ B = \frac{{\cal M}}{\rh} du \wedge dv, ~~~~~~~~~~~~~
A = - 2\frac{e}{\rh}du,  \\
 &&~~~~~~~~~~~~~~~~~~~~~~~  e^{-\Phi}= \sqrt{2 \Lambda} \rh, \nonumber
\eea
which we recognize as a plane fronted wave carrying electric
charge, traveling on the extremal black string of Horne and Horowitz. This
interpretation will be given a thorough justification in the next section,
where we will demonstrate that this solution is in fact directly related to
the extremal limit of our first family by a wave transformation due to
Garfinkle \cite{Gar}.

\section{Traveling Wave on Gauge-Charged Black String}
We begin by briefly reviewing the wave generating technique of
\cite{Gar}.
This technique allows us to superimpose traveling wave contributions
on solutions of Einstein-like theory of gravity with matter couplings of
quite general nature (including stringy gravity) which have null
hypersurface orthogonal Killing vectors. It works as follows. Let
$({\bar g}_{\mu\nu}, B_{\mu\nu}, A_{\mu}, \Phi)$ be a
solution of the equations of motion derived from the action (\ref{w1})
and given in the Einstein frame, with a null hypersurface orthogonal
vector $k$. The Einstein frame is defined by
conformally transforming the world-sheet metric $g_{\mu\nu}$
to ${\bar g}_{\mu\nu}=\exp(-2\Phi)g_{\mu\nu}$.
Then, there exists a scalar field $F$ such that
\be \label{w30}
\nabla_{\mu} k_{\nu} = \frac{1}{2}\bigl(k_{\mu}\nabla_{\nu}F
- k_{\nu}\nabla_{\mu}F\bigr).
\ee
Without changing the matter, we can define the new Einstein frame metric by
\be \label{w31}
{\bar g}'_{\mu\nu} = {\bar g}_{\mu\nu} + F \Psi k_{\mu}k_{\nu},
\ee
where $\Psi$ satisfies
\bea \label{w32}
k^{\mu}\nabla_{\mu}\Psi &=& 0, \nonumber \\
\nabla^{\mu}\nabla_{\mu}\Psi &=& 0.
\eea
We can show that the configuration
$({\bar g}'_{\mu\nu}, B_{\mu\nu}, A_{\mu}, \Phi)$ also represents a
solution of the same equations of motion. The key is to demonstrate
that the equations of motion are invariant under this transformation.
An easy way to see that this is true in the cases we consider is to recall
that the determinant of the metric changes under (\ref{w31}) by a shift
proportional to $k^{\mu}k_{\mu}$. Since $k$ is null, this is zero and
the determinant is invariant: $\det({\bar g}')=\det({\bar g})$. Furthermore,
the field strength of the axion ${\bf H}$ is a three-form, and thus its
dual is invariant under (\ref{w31}), because the metric appears in it only
through the determinant. Additional constraints must be imposed on the
gauge field, however. They are
\bea \label{w33}
k^{\mu}A_{\mu} &=& 0 \nonumber \\
{[ k, A ]} &=& 0
\eea
where the last equation represents the requirement that $k$
Lie-derives the gauge field $A$. These identities then guarantee
the invariance of the gauge field sector under (\ref{w31})
in the equations of motion. In the cases we consider
these identities hold, as we will show below.
Finally, if we compute the Ricci tensors,
which govern the graviton dynamics,
we can show that ${\bar R}'^{\mu}{}_{\nu} - {\bar R}^{\mu}{}_{\nu}$ is
proportional to $\nabla^2 \Psi$, which vanishes by the second condition
of (\ref{w32}). Thus, the Ricci tensor with one contravariant and one
covariant index is also invariant under (\ref{w31}), and so is the Ricci
scalar. This in turn means that all the separate metric-dependent terms
which appear in the equations of motion are invariant, and that (\ref{w31})
indeed represents a motion in the space of solutions, as claimed above.
We remark that the interpretation of the modified solution as a wave
traveling in the original background rests on the property that the
vector $k$ remains a null Killing vector of the final solution too.
This is because the function $\Psi$ is independent of the Killing coordinate
according to the first of the conditions (\ref{w32}). Thus all disturbances
in the metric generated by it must propagate at the speed of light,
without changing its shape.

Now we can apply this technique to the extremal limit of our first family
of solutions. It is evident from the form of
(\ref{w10spec}) that in the extremal
limit $|Q|=m$ the Killing vector $\partial_{x}$ becomes null. For our
purposes it is more convenient to use the shifted Killing coordinate
$\chi=x-\tau/2$, because in the limit when $\sh^2=0$, this solution
reduces to the extremal limit of Horne and Horowitz, and then our
coordinates $\chi, \tau$ are exactly the null coordinates $u,v$.
The Killing vector $\partial_{\chi}$ remains null, as can be seen from
$\partial_{\chi}=\bigl(\partial_{x}\bigr)_{\tau}$.
After the conformal transformation to the Einstein frame, we can rewrite the
solution as
\bea \label{w34}
d{\bar s}^{2} &=& \frac{\rh^2 d\rh^2}{(\rh-m\ch^2)^2}
- 4\Lambda \rh^2 \bigl(1 - \frac{m\ch^2}{\rh}\bigr) d\chi d\tau
+ 2\Lambda m\sh^2 \rh \bigl(1 - \frac{m\ch^2}{\rh}\bigr) d\tau^2, \nonumber \\
    &&~~~ B = \frac{m\ch^2}{\rh} d\chi \wedge dt, ~~~~~~~~~~~~~
A=-\sqrt{2}\frac{m\sh\ch}{\rh}d\tau, \\
    &&~~~~~~~~~~~~~~~~~~~~~~ e^{-\Phi} = \sqrt{2 \Lambda} \rh. \nonumber
\eea
Our null Killing vector is $k=\partial_{\chi}$, with the only nonzero
covariant component $k_{0}=-2\Lambda \rh(\rh - m\ch^2)$.
We can see that the conditions (\ref{w33}) for the gauge field $A$ hold,
since it is directed along $d\tau$ and does not depend on $\chi$.
Therefore, we can apply the wave generating technique. To proceed,
we can see that the scalar $F=1/\rh(\rh - m\ch^2)$ solves the condition
(\ref{w30}). The final step of our calculation consists of finding a
scalar field $\Psi$ which solves the constraints (\ref{w32}). The first
constraint requires $\Psi=\Psi(\rh,\tau)$, and consequently the
d'Alembert operator of the second acquires a particularly simple form
due to this and the properties of the metric. This equation can be written as
\be \label{w35}
\frac{\partial}{\partial\rh} \Bigl((\rh - m\ch^2)^2
\frac{\partial \Psi}{\partial \rh}\Bigr) = 0.
\ee
The most general solution to (\ref{w35}) is
\be \label{w36}
\Psi = g(\tau) + \frac{f(\tau)}{\rh - m\ch^2}.
\ee
The $g$ term here can be dropped because its contribution to (\ref{w31}) is
only a diffeomorphism. The only nonvanishing component of the
matrix $F\Psi k_{\mu}k_{\nu}$ is $F\Psi k_{0}k_{0}=4\Lambda \rh f(\tau)$,
and thus the new Einstein frame metric is
\be \label{w37}
d{\bar s}'^{2}= \frac{\rh^2 d\rh^2}{(\rh-m\ch^2)^2}
- 4\Lambda \rh^2 \bigl(1 - \frac{m\ch^2}{\rh}\bigr) d\chi d\tau
+ 2\Lambda \Bigl(m\sh^2 \rh \bigl(1 - \frac{m\ch^2}{\rh}\bigr)
+ 2\Lambda \rh f(\tau) \Bigr) d\tau^2.
\ee
Consequently, we can rewrite the new solution in the world-sheet frame
by conformally transforming back, using the (unchanged) dilaton, to
get
\bea \label{w38}
ds'^{2}&=& \frac{d\rh^2}{2\Lambda (\rh-m\ch^2)^2}
- 2\bigl(1 - \frac{m\ch^2}{\rh}\bigr) d\chi d\tau
+  \Bigl(\bigl(1 - \frac{m\ch^2}{\rh}\bigr)\frac{m\sh^2}{\rh}
+ \frac{2\Lambda}{\rh}f(\tau) \Bigr) d\tau^2. \nonumber \\
    &&~~~ B = \frac{m\ch^2}{\rh} d\chi \wedge dt, ~~~~~~~~~~~~~
A=-\sqrt{2}\frac{m\sh\ch}{\rh}d\tau, \\
    &&~~~~~~~~~~~~~~~~~~~~~~ e^{-\Phi} = \sqrt{2 \Lambda} \rh. \nonumber
\eea
The wave behavior is completely determined by the function $f(\tau)$.

In the remainder of this section, we will focus on those solutions
(\ref{w38}) where $f={\rm const}$. To start with, we note that if
$f=-m\sh^2/2\Lambda$, the solution is precisely the second extremal
limit of the second family of solutions, discussed at the end of the
previous section. This coincidence reaffirms our choice to label that
solution as a traveling wave on an extremal black string. By the same
token, our starting solution (\ref{w34}) can also be thought of as a
wave of constant amplitude, carrying electric charge and traveling along
an extremal black string. In contrast, no choice of $f$ will lead to
the first extremal limit of the second family of black strings.
We see this because that extremal limit does not have null Killing vectors,
whereas (\ref{w38}) has one.

Another interesting observation related to (\ref{w38}) with $f={\rm const}$
is the effect of this term on the global structure of solutions.
Here the value of $f$ plays the crucial role, dividing the solutions
into three categories, distinguished by the accessibility of the horizon,
located at $\rh_{h} = m\ch^2$, to geodesics probes. If we look at the
geodesic equation for the radial coordinate, with $P_{\chi}$, $P_{\tau}$
the components of the conserved probe momentum in $\chi$ and $\tau$
directions, and $p$ its squared rest mass,
\be \label{w39}
\frac{\dot{\rh}}{2\Lambda} =
\bigl(m\sh^2(\rh - m\ch^2) + 2\Lambda \rh f \bigr)P^2_{\chi}
+2\rh(\rh - m\ch^2) P_{\chi}P_{\tau} - p(\rh - m\ch^2)^2,
\ee
we see that at the horizon, $(\dot {\rh})^2 = 4\Lambda^2 m\ch^2 f P^2_{\chi}$.
Thus, depending on the sign
of $f$, the geodesics either cross the horizon, showing that the proper
extension across it is by including $\rh<\rh_{h}$ (when $f>0$,
and is similar to our first extremal limit of the second family),
cross the horizon but with a mirror-like extension as in our first
extremal limit and the static extremal limit of Horne and Horowitz
(when $f=0$), or bounce back to infinity before reaching the horizon
(when $f<0$, as in our last extremal limit). Therefore, the admissible
global structure is similar either to the
extremal Reissner-Nordstr$\o$m black hole, the extremal static black string,
or the completely nonsingular geometry of our last extremal limit.
In fact, all of these solutions can also be thought of as extremal
black strings, as can be seen from the fact that their  Hawking temperature
is identically zero.

\section{Conclusion}
In this paper, we have presented several new solutions of stringy gravity
in three space-time dimensions. We have found that several of those solutions
admit interpretation as electrically charged black strings, thus generalizing
the static, electrically neutral solutions previously found by Horne and
Horowitz. Our black strings have shown a surprisingly rich geometric structure,
and the existence of several different extremal limits, which are
possible final states which strings can reach by Hawking radiation.
This situation is somewhat akin to that found in the case of gauge-charged
stringy black holes in four dimensions, where the inclusion of the additional
gauge and axion charges has introduced additional dimensions in the space
of allowed parameters describing
a black hole, resulting also in a collection of
new extremal limits \cite{Kall}.
One of our well-defined extremal limits has a null Killing
vector, which we utilized for generating a traveling wave solution on the
string background, employing Garfinkle's techniques. In the special cases
when the wave profile is constant, we have found that the solutions can be
interpreted as yet new extremal black strings, since their associated
Hawking temperature is zero. We have also found that
two of our three extremal limits found directly are mutually related by such
wave transforms.

Our results are highly supportive of the existence of an even more general
family of black objects in three dimensions, which we believe could be
obtained by including an additional
independent parameter, describing the conserved
linear momentum in the direction of the string. Many properties we have
observed indicate this; to name just a few, we could quote multiple extremal
limits, different non-extremal solutions, etc. We think that perhaps the
strongest evidence for this conjecture is our observation of the
apparent inconsistency encountered
in our second black string family. There we have indicated that
on one hand, the presence of the ergosphere should indicate the possibility
of energy extraction via Penrose processes, resulting in the
disappearance of the ergosphere, which, on the other hand, would be
in contradiction with gauge Gauss laws, since the ergosphere in this
case is carried solely by the electric and axion charges. A possible
resolution of this would be the existence of a more general family with
an arbitrary linear momentum along the string, which would be the quantity
to dissipate in the energy extraction by neutral probes. In this scenario,
the string would eventually evolve towards the extremal limit of our
first family, with the gauge charges conserved, but without the ergosphere,
and hence with no possibility of further energy extraction.
We believe that this issue deserves further attention.

\acknowledgments
We would like to thank B. Campbell,
R. Myers, R. Madden, E. Martinez and W. Israel
for helpful discussions. This work has been supported in part by the
National Science and Engineering Research Council of Canada. In addition,
the work of W.G.A. has been supported in part by an Province of Alberta
Graduate
Fellowship, and the work of N.K. has been supported in part by
an NSERC postdoctoral fellowship.


\begin{figure}
\leavevmode
\hbox{\epsfysize=12cm {\epsffile{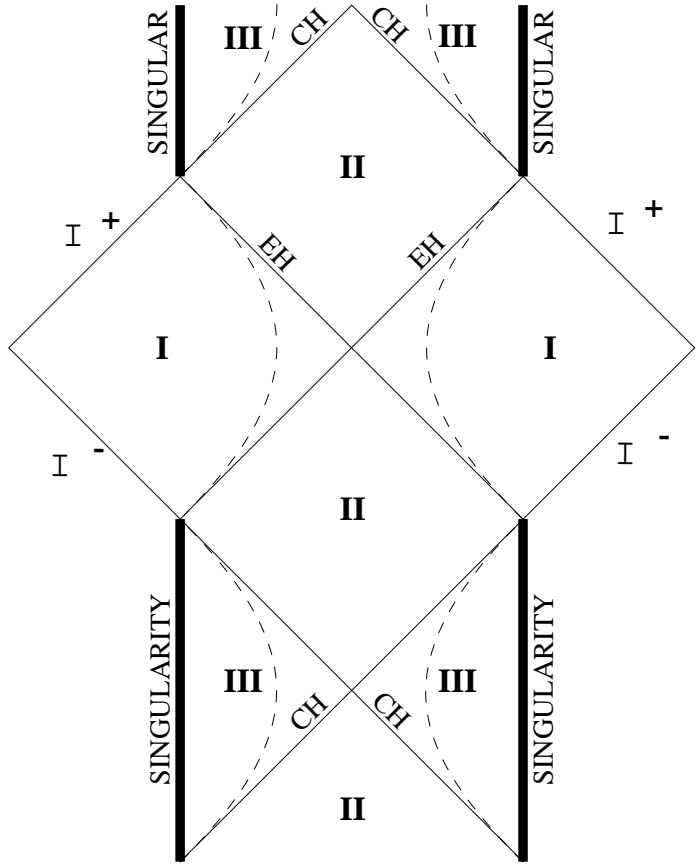}}}
\caption{Causal structure for the two non-extremal black strings (\ref{w10})
and (\ref{w17adm}), as well as for the Reissner-Nordstr$\o$m and
Horne-Horowitz solutions. The hyperbolae denote the static limits
present in our solutions, which do not appear in the previous two cases.
Whereas the static limit inside region III is present in both of
our two cases, the limit in the asymptotically flat region I is the
ergosphere present only in the second solution (\ref{w17adm}).}
\end{figure}

\begin{figure}
\leavevmode
\hbox{\epsfysize=12cm {\epsffile{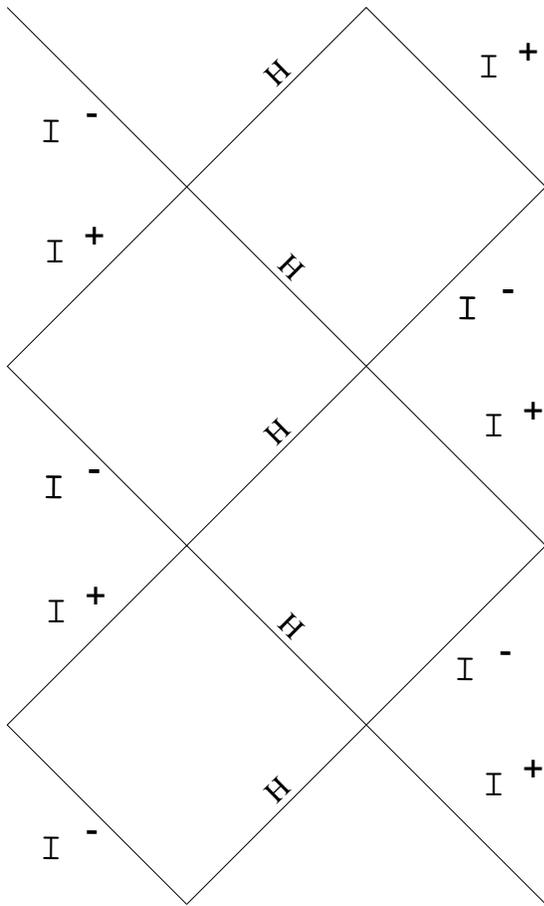}}}
\caption{Causal structure of the extremal limit of the first family
of black strings (\ref{w10}) and the extremal Horne-Horowitz solution.}
\end{figure}

\begin{figure}
\leavevmode
\hbox{\epsfysize=12
cm {\epsffile{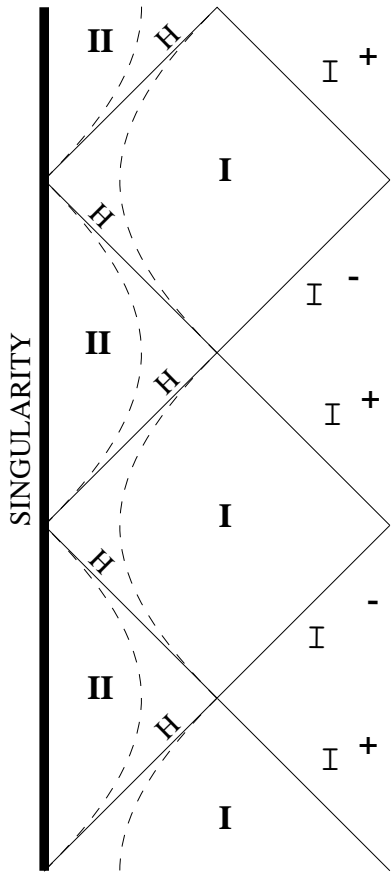}}}
\caption{Causal structure of the $Q^2+4e^2=m^2$ extremal limit of
our second family (\ref{w17adm}) of black string solutions, as well as
of the extremal Reissner-Nordstr$\o$m solution. The hyperbolae here
depict the ergosphere (region I) and the inner static limit (region II)
present in our case, in contrast to Reissner-Nordstr$\o$m.}
\end{figure}

\end{document}